# MINIMISING DELAY AND ENERGY IN ONLINE DYNAMIC FOG SYSTEMS

Faten Alenizi   and Omer Rana

School of Computer Science and Informatics, Cardiff University, Cardiff, UK

*ABSTRACT*

*The increasing use of Internet of Things (IoT) devices generates a greater demand for data transfers and puts increased pressure on networks. Additionally, connectivity to cloud services can be costly and inefficient. Fog computing provides resources in proximity to user devices to overcome these drawbacks. However, optimisation of quality of service (QoS) in IoT applications and the management of fog resources are becoming challenging problems. This paper describes a dynamic online offloading scheme in vehicular traffic applications that require execution of delay-sensitive tasks. This paper proposes a combination of two algorithms: dynamic task scheduling (DTS) and dynamic energy control (DEC) that aim to minimise overall delay, enhance throughput of user tasks and minimise energy consumption at the fog layer while maximising the use of resource-constrained fog nodes. Compared to other schemes, our experimental results show that these algorithms can reduce the delay by up to 80.79% and reduce energy consumption by up to 66.39% in fog nodes. Additionally, this approach enhances task execution throughput by 40.88%.*

## 1. INTRODUCTION

Cloud computing plays an important role in processing tasks generated by IoT devices [1]. However, as the number IoT devices increases, so does the amount of generated data, and processing these data in the cloud may incur significant overhead in multi-hop networks. This is because the cloud server is usually remote and spatially distant from the IoT devices, which leads to high transmission latency, downgraded performance of latency-sensitive applications and network congestion [2]. To address this issue, fog computing (FC) – considered an extension of cloud computing (CC) [3, 4] – has been introduced by CISCO [5] which acts as an intermediary between the cloud and end-user devices. This brings processing, storage, and networking services spatially closer to end devices to reduce latency and network congestion. The main idea behind FC is to deploy computing resources, i.e. fog nodes (FNs), at the network edge [3, 6]. FNs can be routers, gateways, and access points to which end devices offload their computationally intensive tasks. However, FNs are characterised as resource-limited devices [4, 7] because they cannot handle all requests emanating from IoT devices located in their radius of coverage. To overcome this problem, computational offloading within the fog computing paradigm is one of the solutions that helps to improve the utilisation of available resources at FNs and allows for the processing of end-user tasks [8-10].

In fog computing, computational offloading refers to the cooperation between fog nodes in the same layer or with upper layers in which overloaded FNs delegate part of the workload to underloaded FNs within their proximity [9, 11] (we refer to such proximal fog nodes as forming a "neighbourhood"). Offloading, in this sense, refers to the sharing of the workload amongst the nodes to minimise the overall latency of end-user tasks and improve QoS for user applications [11-13]. Computational offloading aims to maximise usage of the available fog resources, however at the expense of increasing energy use of fog nodes. An open research challenge considered in this work is understanding when a FN will decide to dynamically offload its workload and to which other FNs in the neighbourhood. This timely decision making is critical but challenging, especially within a dynamic system, where tasks generation cannot be known a priori. Another important question is understanding how to manage energy of fog resources, while still exploiting most of the available resources. The optimization problem of minimising the delay and the energy consumption has been decomposed into two sub-problems, named delay minimization problem and energy saving problem. The main objectives of this work can be summarised as following: (i) maximising the utilisation of resource-constrained fog nodes; (ii) minimising application loop delay; (iii) improving throughput of user tasks; (v) optimising energy consumption of fog nodes. This paper is structured as follows. The related works are introduced in Section 2. We describe the model of the fog computing system and constraints in Section 3. In Section 4, we explained the problem formulation. In Section 5, the proposed algorithms are presented. Simulations are conducted in Section 6 with numerical results, and concluding remarks are drawn in Section 7 with future work.

## 2. RELATED WORK

This section is divided into two main parts. The first focuses on computational offloading between entities within a specific system model; the second addresses the effect of dynamically controlling servers to improve power efficiency (e.g. either switching servers on and off as required or running them at lower capacity to improve power efficiency).

### 2.1. Computational Offloading

Offloading computing tasks to a cloud data centre and/or neighbouring fog computing servers has received considerable attention in other publications [4, 8, 9, 12-15]. Due to the differences in system models, existing works can be categorised as outlined below:

### 2.1.1. Computation Offloading in IoT-Fog-Cloud Computing Systems

IoT devices can be tasks generators and can process their own tasks aided by either fog or cloud resources, or both. The following scenarios illustrate this point:

(i) An end-user device either processes its tasks locally or chooses to offload them to the nearest fog node for processing (this is referred to as the IoT-Fog computing system). In this scenario, Tang et al. [12] aimed to maximise the total number of executed tasks on IoT devices and fog nodes while meeting the deadline requirements under energy constraints. The authors formulated the problem as a decentralised, partially observable offloading optimisation problem in which end users are partially aware of their local system status which includes the current number of remaining tasks, the level of battery power and the availability of the nearest fog node resource (availability is based on the number of tasks in the fog node's queue). These criteria are used to determine whether to process tasks locally or offload them to the nearest fog node. The suggested solution enables the IoT device to make an approximate optimal decision based on its locally observed system while meeting the delay requirements. Chen and Hao [15] investigated the task offloading problem in a dense software-defined network. The authors describe the problem as a mixed-integer nonlinear problem and decomposed it into two sub-problems: (1) deciding whether the end-user device should process its tasks locally or offload them to the edge device and (2) determining how many computational resources should be given to each task. To solve these sub-problems, the authors proposed an efficient software-defined task offloading scheme. The results of their proposed scheme, compared to random and uniform offloading schemes, demonstrate the effectiveness of their solution in decreasing the overall task execution time and the end-user device's energy consumption.

(ii) To determine whether end-user tasks should be processed on the end-user device, on the fog nodes or on the cloud servers (this is referred to as the IoT-Fog-Cloud computing system), Sun et al. [14] Since processing tasks are not limited to fog nodes, most tasks are processed either on the IoT devices (if they have the computational capacity), or on cloud servers as long as the task deadline is not violated. This leads to fewer tasks being processed within the fog environment, thereby reducing the energy consumed in fog nodes. Computational offloading has been studied in fog radio access networks [13] to achieve minimum system cost, which is the weighted sum of total energy consumption and total offloading latency. In their work, task latency only involves computation and transmission latency – no queuing latency is considered. To enhance the offloading decision, along with improving resource allocation for computation and radio resources, Zhao et al. have formulated the problem as a non-linear, non-convex joint optimisation problem. Their proposed solution has proven its effectiveness in comparison both to the mobile cloud computing system (MCC), in which all user tasks are processed on a cloud server, and to the mobile edge computing system (MEC), in which all end-user tasks are executed in the edge computing system. This is attributed to the fact that, in their model, both the fog and the cloud computation resources are available to support the offloading scheme.

(iii) In this scenario, in addition to the previous scenario, fog nodes can make the decision to offload the workload partially or fully to another fog node (this is referred to as the IoT-Fog-Fog-Cloud computing system. In [11], Yousefpour et al. proposed a delay-minimisation policy to reduce the overall service delay. In their work, the decision for a fog node to process its upcoming task(s), or either to offload to one of its neighbours (horizontal cooperation) or to the cloud server, is based on the estimated queue waiting time. If the offloading (queue waiting time) threshold has been reached, then a fog node will select one of the neighbouring fog nodes in its domain to offload its upcoming tasks to. The selection of the best neighbouring fog node is based on minimising total propagation delay plus queuing delay. Three different models have been considered and compared. In the first model, there is no processing of tasks in the fog system (NFP), so an IoT device either processes its own requests or sends them to a cloud data centre. The second model only allows for the processing of light computational tasks in the fog system with heavy computational tasks being processed in the cloud (LFP). In the third model, fog computing can process all types of tasks (AFP). The results show that AFP achieved the minimum average service delay compared to the two previous models.

**2.1.2. Computation Offloading in Fog-cloud Computing Systems**

In this approach the end-user devices offload all their computational tasks to the associated fog nodes for processing. The associated fog nodes can choose to offload part of their computational workload to another fog node or to the cloud, thereby exploiting fog and cloud resources to process end-user tasks. Gao et al. [9] investigated dynamic offloading and resource allocation, formulating the problem as a stochastic network optimisation to minimise delay and power consumption while ensuring the stability of all queues in the system. They present a predictive offloading and resource allocation approach that focuses on the trade-off between energy consumption and delay. Their approach suggests that increasing the allocation of processing resources in fog nodes causes a reduction in delay but increases energy consumption due to the processing of additional tasks and vice versa. The authors demonstrate the benefit of their approach compared to other schemes.

In [4], Xiao and Krunz developed a workload offloading strategy that maximises the average response time of all end-user tasks that are given a power efficiency constraint. In their experiment, power consumption is measured as the power spent on offloading each unit of received workload, but the power consumed to execute workloads is not considered. The decision for fog nodes to start cooperating and offload the workload is made through an agreement between the parties, and the amount of workload to be offloaded is based on the workload processing capabilities and the workload arrival rates. Based on their results, cooperation between fog nodes helps to decrease the average response time. They also observed a fundamental trade-off between the average response time and the power efficiency of the fog node. The authors suggested that in order to optimise the power efficiency of the fog computing systems the response time of end-user tasks should be set to its maximum tolerable point, which means that when end-user tasks can stand higher delay there is no need to offload tasks to save energy. In addition, the authors stated that with delay-sensitive applications it is better to equip fog nodes with high-power consumption so that they are able to share more of their workload with other fog nodes, thus minimising response time.

### 2.1.3. Computation offloading in Fog Computing Systems

In this computing system, only fog resources are available to process end-user tasks, but no processing in the cloud is considered. Considering computation resources, Mukherjee et al. [16] designed a scheduling policy that manages to meet the deadline constraint of end-user tasks. In their scheduling policy, the deadline requirements of a task determine whether a fog node places it in its high priority queue, in its low priority queue or offloads it to one of its neighbouring fog nodes within the same tier. The decision whether to process the task or offload it to its neighbours is based on the availability of a neighbour with a lower transmission delay plus lower queue length of a specified type. Their results show the effectiveness of their proposed policy when compared to (a) an approach where no offloading is involved, and the fog node assigns its upcoming tasks randomly to one of its two queues with no consideration to priority and (b) an approach where workload offloading occurs between fog nodes with random task scheduling to any queue without considering their priority.

## 2.2. Dynamic Server Energy Management

Related work in this area is classified into two sections based on the environment in which this technique has been applied, which are Cloud Computing systems and Wireless Local Area Networks (WLANs). This technique has not been applied in fog computing system, while it has proved its efficiency in other environments.

### 2.2.1. Cloud Computing System

To save energy in a cloud environment, it has been proposed that servers should be dynamically shut down [17, 18] or put into sleep mode [19-21]. The authors in [17-21] investigated the problem of Virtual Machine (VM) placement to save energy and still maintaining QoS. In their work, underloaded data centres were detected and shutdown as per [17, 18] or put in a sleep mode as per [19-21]. All VMs in those data centres where then be migrated to other active underloaded data centres. This is done to minimise the energy consumed by cloud computing systems and is called 'VM consolidation'. For overloaded data centres, different VM selection methods have been proposed to decide which VMs should be migrated to other active data centres. Furthermore, a switched-off data centre could be activated to accommodate the migrated VMs to ensure QoS requirements in the system are met. The researchers saved the most energy when putting idle-mode data centres into sleep or shutdown mode.

Mahadevamangalam in [19] stated that in a cloud environment, idle-mode data centres with no workload consume energy equivalent to 70% of the energy consumed by data centres that are fully utilised and in busy mode. Therefore, shutting down idle-mode data centres will save up to 70% of the energy consumed in a cloud environment.

### 2.2.2. Wireless Local Area Networks (WLANs)

In WLANs, putting access points (APs) into sleep mode or switching them off has improved the energy efficiency of WLANs. In [22], Marsan and Meo found that in a group of APs that partially overlap, having one AP in each group to monitor the system and serve the upcoming users while all others are switched off can reduce energy consumption by up to 40%. In addition, if all APs are switched off during idle periods, e.g. at night, energy consumption could be reduced by a further 60%. Li et al. [23] proposed a state transitions-aware energy-saving mechanism in which APs are not just switched on and off based on user demand, but there is also an intermediate stage that helps make the switching frequency as low as possible. This is to avoid frequently switching APs on and off as this will shorten their service life and also to avoid latency and energy overheads when APs are switched on.

## 2.3. Summary

In connection with minimising delay, computational offloading has been proposed in the literature [4, 8, 9, 11-16, 24]. Computational offloading can be deployed offline or online. In offline deployment, computational offloading decisions are made at the system design stage. All the required information about the system is known beforehand and is based on historical or predictive knowledge, such as the computational capacity of fog nodes, the total number of IoT devices and their workload (number of requests). In online deployment, the decision of computational offloading takes place at run-time and considers the current system status and process characteristics, such as the current waiting time. Most research investigating the offloading problem have considered the offline approach [4, 8, 9, 14, 15, 25] , while online approach has limited coverage [11-13, 16]. This shed lights on the importance of investigating the online computational offloading method, our approach primarily makes use of the online approach. In addition to that, none of the state-of-the-art fog computing models explore the impact of varying the offloading threshold on the system performance.

Fog computing is designed to place computational resources near the end users. To minimise delay, the end-users send their requests to the nearest fog node. However, if the fog node is overloaded, existing mechanisms focus on offloading part of this workload to the cloud for processing. However, there might be other nearby underloaded fog nodes that could help to process the workload to further minimise delay. This is called 'fog cooperation', but so far it has received limited coverage [4, 9, 11, 16].

In terms of minimising both delay and energy in the fog paradigm, most studies have addressed either minimising energy at IoT devices and ignoring the energy spent at the fog paradigm [12, 13, 15] or investigating the trade-off between these two aspects withing fog systems [4, 9, 14] because executing more tasks at fog nodes will reduce delay and consume more energy while executing fewer tasks at fog nodes and sending the rest to the cloud will increases delay but reduces energy consumption at fog paradigm. Therefore, most research addresses the balance between delay and energy by processing the workload on IoT devices, fog nodes or cloud servers if the QoS is satisfied. This results in fewer tasks being processed by the fog, thus consuming less energy as long as the QoS is met (e.g. deadline of users' tasks). However, there might be underloaded or idle-mode fog nodes that could be switched off to save energy but still maintain the benefits of fog architecture, i.e. executing more tasks at fog nodes and thus minimising delay. To the best of knowledge of this paper's authors, minimising both delay and energy at the same time and applying dynamic server energy management by switching on/off fog nodes have not been addressed before in the fog system.

## 3. SYSTEM MODELLING AND CONSTRAINTS

System model is presented in section 3.1, and Types of Connections and Constraints is described in section 3.2.

### 3.1. System Model

Network diagram is described in section 3.1.1, and application module description in section 3.1.2.

**3.1.1 Network Diagram**

An overview of the fog computing architecture is shown in Figure.1 and consists of three layers:

- **The IoT devices layer:** this layer contains of mobile vehicles. The vehicle node has a set of sensors. Each sensor transmits different types of tasks and an actuator and once they are within the coverage radius of a fog node, they will send their tasks. Two types of tasks are emitted by the mobile vehicle. The first type is non-urgent and contains information such as current location, speed, and direction of the vehicle. The second task is an urgent request that requires a quick response. For example, this task may contain a video stream of a moving vehicle's surroundings, which requires quick processing by fog nodes to help avoid collisions. This might be important, especially for autonomous driverless vehicles.
- **Fog computing layer:** this layer consists of a set of fog nodes and a fog controller. Fog nodes reside in roadside units (RSU) that are deployed in different areas of a city. Fog nodes can communicate with each other if they are located within each other's vicinity [26]. Fog nodes can form an ad hoc network between themselves to share and exchange data. All fog nodes are logically connected to the fog controller which monitors the performance of all fog nods and manages the resources. The fog nodes are static and receive two different types of tasks from all vehicles within their radius. These tasks are called priority and non-priority tasks. Regarding priority tasks, fog nodes process requests generated by a user's sensor and send the response back to the user. For non-priority tasks, fog nodes do some processing of the information provided by the vehicles within their range and send the results to the cloud for further analysis and storage for retrieval by traffic management organisations.
- **Cloud computing layer:** this layer contains cloud servers. It manages and controls the traffic at the city-level based on historical data received by fog nodes.

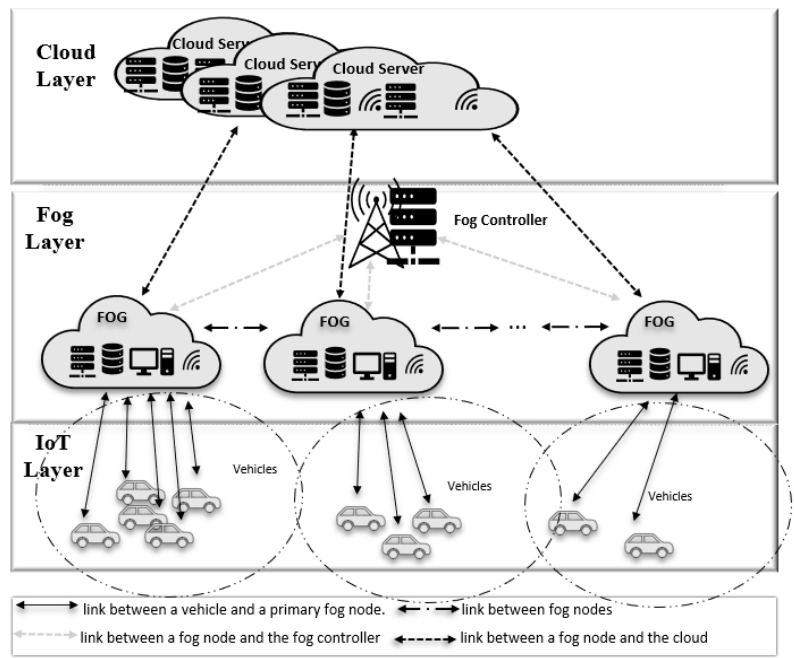

Figure 1: Fog Computing Model

**3.1.2 Application module description**

The application model of this study consists of three modules named Road Monitor, Global Road Monitor and Process Priority Tasks. The first two modules are responsible for traffic light control systems and the last module is only for processing end-user priority tasks. The function of each of these modules is as follows:

• **Road Monito**r: this module is placed in fog nodes. If a vehicle enters an area within the coverage of a fog node, the sensor automatically sends the current car location, its speed, weather conditions and road conditions to the connected fog node for analysis. Then the module processes these data and the results are sent to the cloud for further analysis.

• **Global Monitor**: this module is placed in the cloud and receives the collected data from fog nodes (after being processed by the Road Monitor module), analyses these data and stores the results.

• **Process Priority task**: this module is placed in fog nodes and is responsible for processing the priority requests from the user. The results are then sent back to the user. The application in iFogSim is represented as a directed acyclic graph (DAG) = (M, E) where M is the set of application modules deployed = {m1, m2, m3, ... , mn}, e.g. Process Priority Task, Road Monitor and Global Road Monitor modules. Between application modules, there is a set of edges belonging to E, which represents the data dependencies between application modules. This is shown in Figure.2.

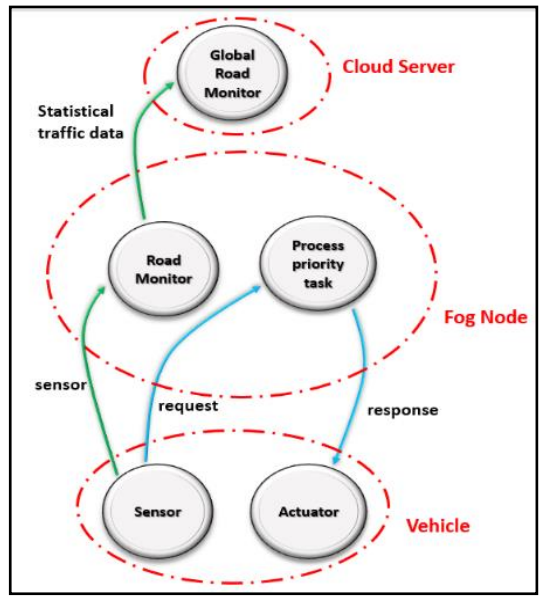

Figure 2: Directed Acyclic Graph (DAG) of the application model.

## 3.2 Types of Connections and Constraints

This section describes the connections between a vehicle and a fog node, between fog nodes, and between fog nodes and cloud. Also, the set of constraints involved within these connections.

### 3.2.1 Connection between Vehicles and Fog nodes

The connection between a vehicle and a fog node is made with communication and processing constraints.

- **Communication Constraints**

Vehicles connect to the fog node if and only if it is located within its communication range, as constraint (1)

$$D_{v,f} \leq max\ Coverage_f;\ \forall\ v \in V, \forall\ f \in FN \qquad (1)$$

Where V is all vehicles, v one vehicle, FN is all fog nodes and f is one fog node. $D_{v,f}$ is the distance between a vehicle v and a fog node f, is calculated as

$$D_{v,f} = \sqrt{(X_v - X_f) + (Y_v - Y_f)}; \quad \forall v \in V, \forall f \in FN \tag{2}$$

where $(X_V, Y_V)$ and $(X_f, Y_f)$ are the coordinates of a vehicle and a fog node, respectively. If a vehicle is located within the coverage radius of more than one fog node it will connect to the nearest fog node. This to reduce delay because the expected arrival time of the task at the connected fog node depends on the transmission and the propagation delay, but the propagation delay depends solely on the distance between the two connected objects. Propagation delay (PD) is calculated as

$$PD = \frac{D_{v,f}}{PS} \tag{3}$$

Following [27], we assume that the speed of signal propagation (PS) is equal to the speed of light, $c = 3 \times 10^8$.

- **Processing Constraints**

For fog nodes to process user tasks, application modules in which these tasks are processed should be placed at fog nodes. To ensure the placement of these application modules, application modules require CPU, Ram and Bandwidth capacity so that fog nodes will have enough CPU, Ram and Bandwidth capacity to place these application modules, thus processing end-user tasks at the fog paradigm.

$$\sum_{i=0}^{M} Required_{Capacity}\, m_i \leq \sum Available_{Capacity}\, f; \quad \forall m_i \in M, \forall f \in FN \tag{4}$$

Where $Required_{Capacity}$ for each application module = {CPU, Ram, Bandwidth} and the fog node capacity = {CPU, Ram, Bandwidth}. Constraint (4) ensures that the total required capacity of all application modules should not exceed the available capacity of the fog node in which they should be placed. In iFogSim, if the capacity required to place application modules exceeds the available capacity of fog nodes, the system will iterate through upper tiers fog computing system until it reaches the cloud and places these application modules. The CPU required for an application module is calculated as following:

$$CPU = NV * (Rate * TaskCPU) \tag{5}$$

Where NV is the total number of connected vehicles to a fog node, and TaskCPU is the task CPU length which is the number of instructions contained in each task in Million Instructions Per Second (MIPS). Rate is calculated as:

$$Rate = \frac{1}{Transmission\ Time\ in\ ms} \tag{6}$$

The placement of application modules in iFogSim is done before running the system and starting the emission of tasks. If the number of vehicles increases, this will impact the required CPU capacity for an application module. In this case the number of connected vehicles for each fog node is limited as constraint (7).

$$\sum_{i=0}^{V} v_i f_j \leq MAX_{vehicle\ number}; \quad \forall v_i \in V, \forall f_j \in FN \tag{7}$$

**3.2.2 Connection between Fog nodes**

This section describes the waiting queue for fog nodes in which the offloading decision is determined, how fog nodes communicate and the selection criteria for the best neighbouring fog node.

- **Fog nodes' waiting queue**

Each fog node maintains a waiting queue into which tasks are placed upon their arrival at the fog node. Fog nodes process one task at a time. Once the execution of that task is completed the fog node will check its waiting queue and process the next task according to its scheduling policy, i.e. first come, first served. This process continues until no tasks are in the waiting queue. Following the work [11], the waiting queue time triggers the decision to start computational offloading to neighbouring fog nodes. To start sharing workloads, the queue waiting time ($T^{Queue}$) should exceed the offloading threshold, e.g. 50ms, 100ms or 200ms.

$$T^{Queue} > Max_{threshold} \qquad (8)$$

$T^{Queue}$ is calculated as

$$T^{Queue} = \sum T_i * T_i^{process} + \sum T_z * T_z^{process}; \forall\, i, z \in T \qquad (9)$$

Where $T_i$ and $T_z$ are the total number of tasks of the type *i* and *z*, e.g. priority or non-priority. T is all tasks and $T_i^{process}$ is the expected execution time of a specific task and calculated as

$$T^{process} = \frac{TaskCPU}{F\_MIPS * N\, of\, PS} \qquad (10)$$

Where F_MIPS is the total mips available in a fog node and N of PS is the total number of processing units allocated in that fog node.

- **Coverage Method**

To achieve area coverage, several fog nodes are required. Fogs can also overlap to achieve maximum coverage as in [28] see Figure. 3.

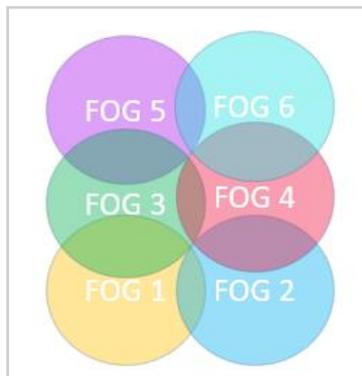

Figure 3: Overlapping Fog Nodes.

- **Selecting the Best Neighbouring Fog Node**

The process of selecting the best neighbouring fog node follows the work in [11]. It happens when a fog node reaches its offloading threshold, e.g. 50ms, 100ms or 200ms waiting queue time, for each upcoming task that is generated from vehicles in the coverage range of this fog node.

The neighbouring fog nodes of a fog node are the fog nodes that are located within the coverage radius of the fog node itself. This is shown in constraint (11)

$$d_{ij} \leq Coverage_{radius}; \quad \forall \, i,j \in FN \tag{11}$$

Where $d_{ij}$ is the distance between fog nodes i and j. In Figure. 3, FOG 2 and FOG 3 are the neighbouring fog nodes for FOG 1. Also, FOG 1, FOG 4, FOG 5 are the neighbouring fog nodes for FOG 3. The criteria for selecting the best neighbouring fog node depends on two factors. First, the neighbouring fog node should be within the communication range of the primary fog node. Second, and most importantly, a neighbouring fog node should have the minimum sum of waiting queue time plus propagation delay amongst all available neighbours.

$$Min \sum T^{Queue} + PD \tag{12}$$

PD is calculated as

$$PD = \frac{D_{f,f'}}{PS} \tag{13}$$

($D_{f,f'}$) is the distance between fog nodes f and f' and it is calculated similar the distance between a vehicle and a fog node ($D_{v,f}$) and propagation speed PS is equal to the speed of light, its value $3 \times 10^8$, this done similar to the work in [27].

### 3.2.3 Between Fog Nodes and the Cloud

When fog nodes finish the processing of non-urgent tasks the results are sent to the cloud for further analysis and processing by the application module named Global Road Monitor. In the current work, the cloud is the least to be considered in sharing the workload of fog nodes when they reach the offloading threshold. This is due to the availability of neighbouring fog nodes and in order to get maximum usage of the available resources in the fog system. However, if all neighbours reach their offloading threshold, the primary fog node will determine to send the task to the cloud if its queue waiting time is higher than transmission delay caused by sending the task for processing to the cloud and getting the results back. Due to the powerful computational capabilities at the cloud server compared to fog nodes, queueing delay is neglected so tasks are processed upon their arrival [29-31] .

## 4. PROBLEM FORMULATION

The optimisation problem of minimising the delay and the energy consumption has been decomposed into two sub-problems: the delay minimisation problem and the energy saving problem.

### 4.1 Delay Minimization Problem

The response time includes the round-trip time for transmitting the workload between a user and the associated fog node. It includes the transmission delay, propagation delay, queuing delay and processing delay. If the workload is processed by the vehicle's primary fog node then the service latency is calculated as

$$T = T^{sTv} + 2 X (T^{Transmission}_{vTf} + PD_{vTf}) + T^{Queue} + T^{procss} + T^{vTa} \tag{14}$$

Where $T^{sTv}$ and $T^{vTa}$ is the latency time between a vehicle and its sensor, and between the vehicle and its actuator, respectively. $T^{Transmission}_{vTf}$ is transmission delay between the vehicle and its primary fog node. It is based on the network length of the task and the bandwidth, and it is calculated as

$$T^{Transmision} = \frac{Network\ Length\ of\ Task}{Bandwidth} \quad (15)$$

If the primary fog node decides to offload the workload to one of its neighbours, then the latency is calculated as

$$T = T^{sTv} + 2x(T^{Transmsiion}_{vTf} + PD_{vTf}) + 2x(T^{Transmission}_{fTf} + PD_{fTf}) + T^{Queue} + T^{Process} + T^{vTa} \quad (16)$$

If the primary fog node decides to send the task to the cloud, then the latency is calculated as

$$T = T^{sTv} + 2x(T^{Transmsiion}_{vTf} + PD_{vTf}) + 2x(T^{Transmission}_{fTc}) + T^{Process} + T^{vTa} \quad (17)$$

### 4.2 Energy Saving Problem

By minimising the power consumption of fog nodes, the overall cost of electricity consumption and environmental impact is reduced. Each fog node has two power modes: idle and busy. The fog node's power is said to be in idle mode when the fog node is not doing any task processing and in busy mode when the fog node is busy processing tasks. The energy consumed is the power spent when a fog node is processing workload and when the fog node is switched ON and not doing any processing. The total energy consumption in iFogSim is calculated as in [32] as

$$E = PR + (TN - LUT) * LUP \quad (18)$$

Where PR is previous total energy consumed in this fog node, TN is the time now which is the time that the updateEnergyConsumption () is called when utilising this fog node, LUT is the last time this fog node has been utilised and finally LUP which is the fog node last utilization power status, which is idle power or busy power, the value of this is based on the predefined parameters when creating a fog node. The problem of minimizing delay and energy is formulated as follows:

$$\text{Min} \sum T \ \& \ \sum E$$

$$\text{s.t.} \ (1), (7), (4)$$

$$T^{Queue} \leq Max_{threshold} \quad (19)$$

$$P_F + P_N = 1, P_F \ \& \ P_N = \{0, 1\} \quad (20)$$

Equation (1) ensures the connection between a fog node and a vehicle that is located within its communication range. Equation (7) ensures the number of vehicles connected to one fog node does not exceed the threshold number. Constraint (4) ensures the placement of application modules at fog nodes. Equation (19) ensures the stability of fog nodes' queues so that, to process its upcoming tasks, the waiting queue time should not exceed its threshold. In constraint (20), PF and PN mean that if the task is processed in its primary fog node, then PF = 1 and PN = 0 and vice versa. Therefore, the task is either processed in the primary fog node or one of its neighbours.

## 5. PROPOSED ALGORITHMS

An approach that combines two algorithms has been proposed to solve the above stated problem. The first algorithm is called dynamic task allocation and the second is called dynamic resource saving. In this paper, both stated algorithms need to work together to achieve the intended outcome.

## 5.1 Dynamic Task Scheduling (DTS):

The aim of this algorithm is to minimise delay by allowing cooperation between fog nodes in terms of workload sharing, to maximise the resource utilization and maximise throughput. The fog controller is not involved in the selection of the best neighbouring fog node, it is mainly involved in the DEC algorithm. Also, in regards to DTS algorithm, if the best neighbour is switched OFF, the fog controller will send a signal to switch ON the selected best neighbour, this is further explained in section 5.2.

The process of offloading a task based on the queue waiting time of the fog nodes was originally proposed by [11]. In [11], the task can be offloaded multiple times, which means that if the primary fog node decides to offload the upcoming task to its neighbour $i$, by the time this task arrives at fog node $i$, fog node $i$ might have reached its offloading threshold. Then fog node $i$ will select fog node $j$ to offload this task to, resulting in offloading this task multiple times and adding additional transmission and propagation delay. As stated by [11], multiple task offloading will increase the delay compared to only allowing the task to be offloaded one time, and this is applied to the current work. The technique is shown in Figure. 4.

When a fog node receives a task, if this task is the first task in its queue it will immediately process it, if not, it will check its queue waiting time. If its queue did not reach its offloading threshold, e.g. 50ms, 100ms or 200ms, the task will be added to its queue, but if the queue reaches its threshold the fog node will check if the task has been offloaded by another fog node. If it has, then it will add this task to its queue. If it has not been offloaded by another fog node it will select the best neighbour to offload this task to, according to the criteria described in section 3.2.2. If the best neighbour reaches its offloading threshold during the selection process and before offloading the task, then the primary fog node will make the decision whether to offload the task to the cloud for processing or process the task locally. This is determined when comparing the transmission delay caused by sending the task for processing to the cloud and getting the results back with the queuing delay of the fog node itself. if the queueing delay is higher, then the fog node will send the task for processing to the cloud, else, the task will be processed locally at the primary fog node.

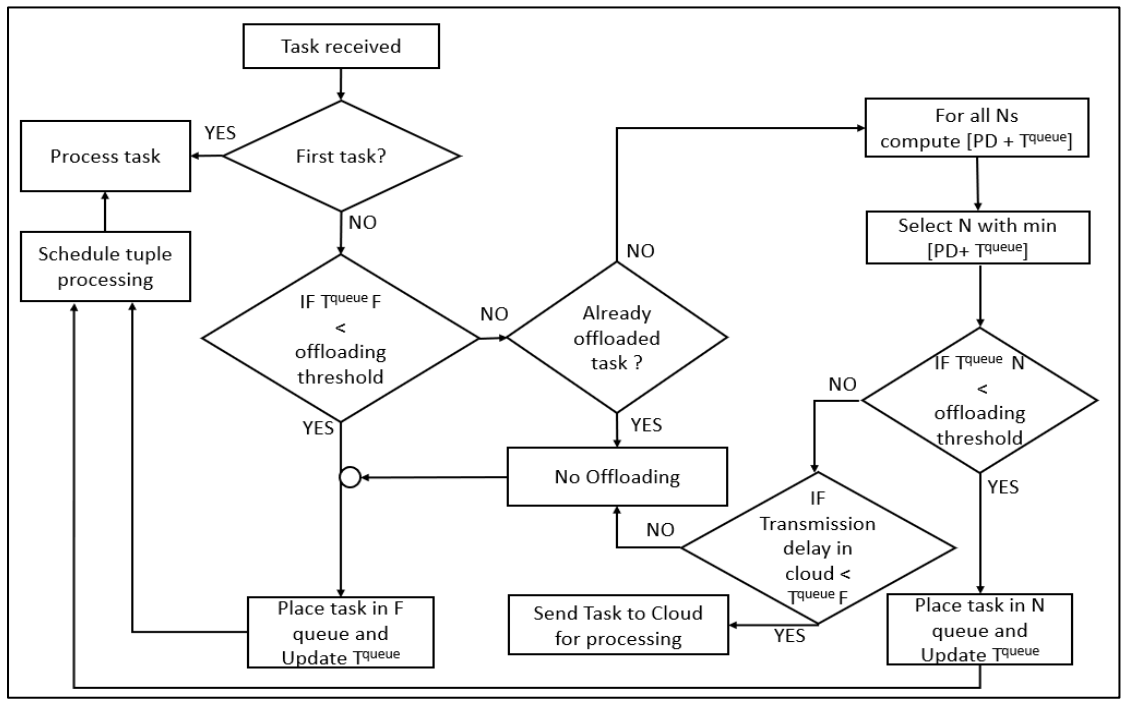

Figure 4: Flowchart of Dynamic Task Scheduling Algorithm.

### 5.2 Dynamic Energy Control (DEC)

The need for 24/7 availability of fog nodes poses a challenge on energy efficiency and cost since the fog provider needs to maintain available resources that may be used but are not continuously needed. If a fog node is not needed it should be turned off to save energy. Dynamic energy control (DEC) has been proposed in order to optimise resource utilisation by dynamically deciding when to switch off an active fog node(s) and conserve overall system energy. The pseudo code of our proposed algorithm is given in Algorithm 1.

In this system, the ON and OFF switching of fog nodes is carried out by the fog controller which runs algorithm 1 each time it receives information about the system. Fog nodes update the fog controller with their information so that fog controller can make the appropriate decision to save energy. At the beginning of the simulation, all fog nodes are switched OFF.

---

**Algorithm 1** Dynamic Energy Controlling

**Input:** System Data: 1- current waiting time; 2- current processing states; 3- if awaiting task/s
**Output:** Sending signals to switch ON/OFF determined FNs

1:   Fog Controller receives System data
2:   **for** all FNs **do**:
3:     **if** (FN. status ==**OFF**)
4:       **if** (FNQueueSize! = 0)
5:         **Send** Signal **ON**
6:       **else**
7:     **else**
8:       **if** (processingStatus =1) //fog node is not processing task/s
9:         **Send** Signal **OFF**
10:       **else**
11:     **end if**
12:   **end for**

---

## 6. PERFORMANCE EVALUATION

In this section, we first provide the details of the simulations, then we investigate the performance of our two comined algorithms.

### 6.1 Simulation Environment Settings

iFogSim has been used to simulate the environment. It is a toolkit developed by Gupta et. al [33], which is an extension of the CloudSim simulator. It is a toolkit allowing the modelling and simulation of IoT and fog environments and is capable of monitoring various performance parameters, such as energy consumption, latency, response time, cost, etc. For this research, the three-tier fog system was established first as shown by the simulation in Figure. 1. The simulation was run with one cloud server, seven fog nodes, the fog controller, and a total of 50 vehicles. Two fog nodes connected to 25 vehicles, but the other five fog nodes are not connected to any vehicles. This is done to vary the workload amongst fog nodes because if all fog nodes have the same workload then offloading will not be beneficial [11]. Each vehicle transmits two different tasks every 3ms. The parameter values used in the simulation is in Tables 1 - 5.

Table 1: Application Modules Requirements.

| Module | CPU (mips/vehicle) | BW (Mbps) | Ram (GB) |
|---|---|---|---|
| Process priority task | 333.33 | 1000 | 10 |
| Road Monitor | 300 | 1000 | 10 |
| Global Road Monitor | 99.99 | 1000 | 10 |

Table 2: Tasks details.

| Task Type | Processed module | CPU length (MIPS) | Network Length (Mbps) |
|---|---|---|---|
| Request (urgent) | Process priority task | 1000 | 1000 |
| Sensor (nonurgent) | Road Monitor | 900 | 500 |
| Statistical traffic data | Global Road Monitor | 300 | 500 |

Table 3: Entity Configurations in iFogSim.

| Characteristics | Vehicle | Fog nodes | Cloud servers |
|---|---|---|---|
| CPU (MIPS) | 0.0 | 15100 | 448000 |
| RAM (MB) | 0 | 40000 | 40000 |
| Uplink BW (Mbps) | 1000 | 1000000 | 1000000 |
| Downlink BW (Mbps) | 1000 | 1000000 | 1000000 |
| Rate Per MIPS | 0.0 | 0.001 | 0.01 |
| Level | 2 | 1 | 0 |

Table 4: Power Consumption with ON/OFF.

| Device | Power Consumption (W) when device is ON | | Power Consumption (W) when device is OFF |
|---|---|---|---|
| | Idle | Busy | Power |
| Fog Node | 83.4333 | 107.339 | 0.0 |
| Cloud Server | 16*103 | 16*83.25 | No |

Table 5: Latency values between entities.

| Between | | Link latency (ms) |
|---|---|---|
| Cloud | Fog node | 100 ms |
| Fog node | Neighboring FN | 2 ms |
| Vehicle | Fog node | [1-5] depends on location |
| Sensor/Actuator | Vehicle | 1 ms |

## 6.2 Experiments

The conducted experiments are shown in Table 6. The metrices used to measure the performance are:

- **Service latency** as the average round trip time for all tasks processed in the fog environment
- **Throughput**, which is measured as the total number of processed tasks within a time window.
- **Total Energy Consumption** in fog environment

Table 6: Set of Conducted Experiments Details.

| Experiment | | Dynamic Task Scheduling | | Dynamic Energy Controlling |
|---|---|---|---|---|
| no | name | Yes/No | When | |
| 1 | No offloading | no | - | No |
| 2 | | no | - | Yes |
| 3 | Offloading-50 | yes | 50 ms | No |
| 4 | | yes | 50 ms | Yes |
| 5 | Offloading-100 | yes | 100 ms | No |
| 6 | | yes | 100 ms | Yes |
| 7 | Offloading-200 | yes | 200 ms | No |
| 8 | | yes | 200 ms | Yes |

**6.2.1 Average round trip time**

There are two control loops in the simulation:

• Sensor → Process Priority Tasks → Actuator. This control loop represents the path of the priority requests, and it is called Control loop A.

• Sensor → Road Monitor → Global Road Monitor. This control loop represents the path pf the non-priority requests, and it is called Control loop B.

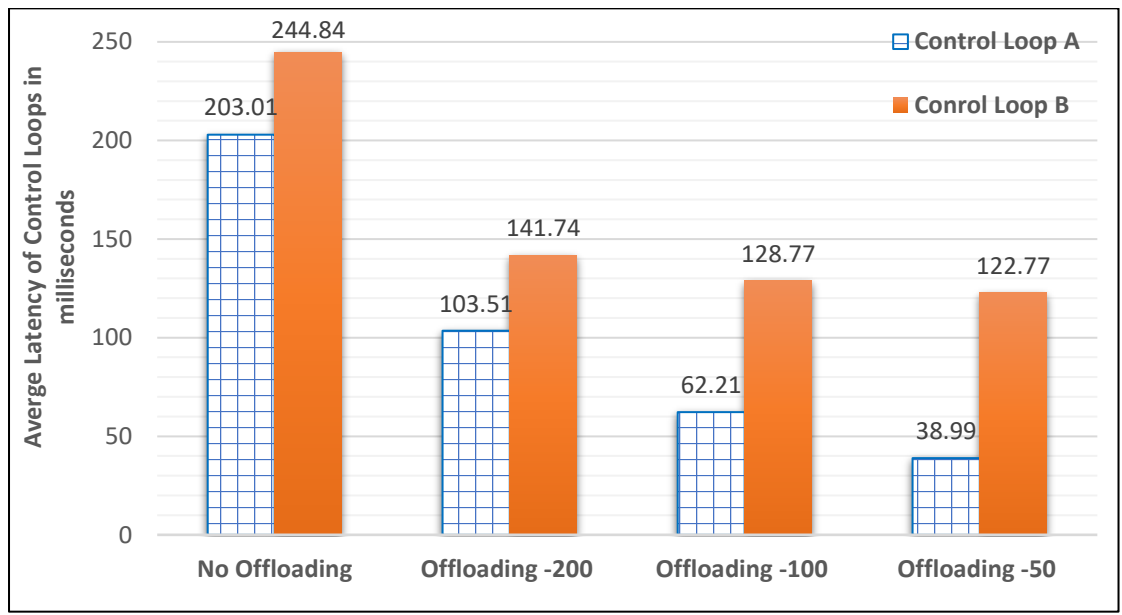

Figure 5: Average Round Trip Time with no-offloading and different Offloading Thresholds

The aim here is to minimise the average round-trip time for control loop A, in which the result is going back to the users, compared to control loop B, in which the user tasks should be processed at fog nodes and the results sent to the cloud for further analysis and storage. The results in Figure. 6 show that when a fog node is not offloading its tasks to the neighbouring fog nodes, the average round trip time for all the processed tasks for control loop A is 203.01ms. This is due to the long queuing delay. However, the average round trip is minimised when the offloading threshold is set to 50ms. This is because more neighbours are involved in the process of executing tasks. With a 50ms threshold, the average latency of the control loop was reduced by 80.79% compared to the no-offloading case.

### 6.2.2 Throughput Evaluation

According to the results in Figure. 7, when the offloading threshold is set to 50ms the number of executed tasks is increased by almost 40.88% compared to the no-offloading method. As with the no-offloading method, many tasks are waiting to be executed in the queue compared to when the offloading threshold is set to 50ms, the threshold where fog node cooperation is allowed and workloads (tasks) are shared.

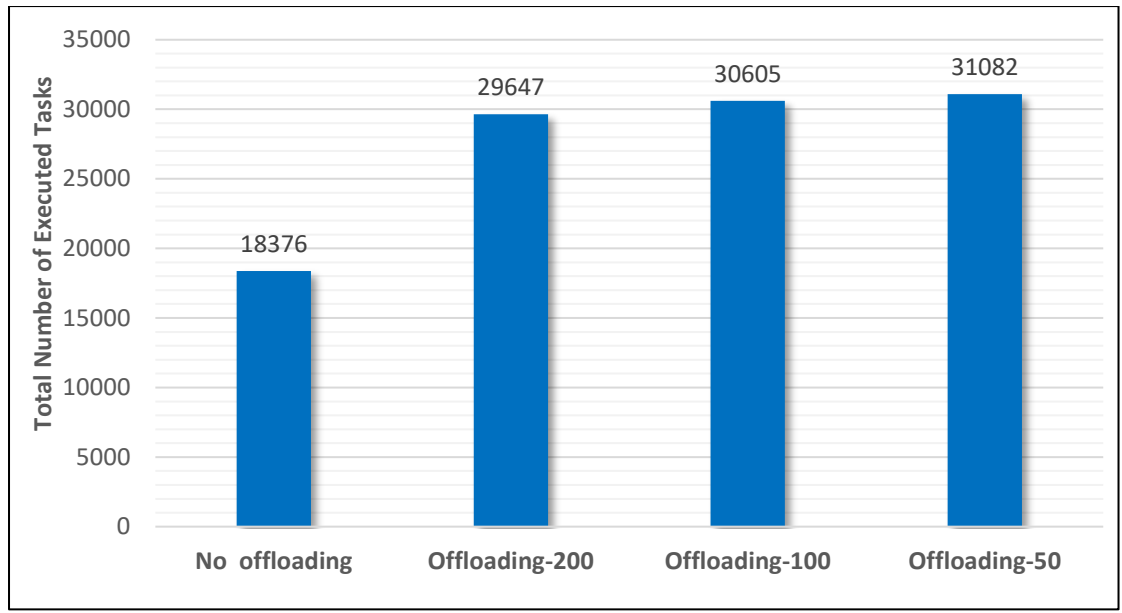

Figure 6: Number of executed Tasks in Fog Nodes with no offloading and different Offloading Thresholds

### 6.2.3 Total Energy Consumption

In cases where a dynamic energy control algorithm is not applied the highest energy consumption in the fog environment occurs when the offloading threshold is set to 50ms. This is because more fog nodes are involved in the execution process and are in their busy mode power mode. This compares to the no-offloading method where only two fog nodes are busy processing tasks while the rest of the fog nodes are not doing any processing and are in their idle power mode (see Figure 8). In the no-offloading method, DEC saves around 66.39% of power. This power was spent powering on unused fog nodes, which cause a wastage in resources. Applying the DEC algorithm helps to minimise the total energy consumed in the fog environment by 2.59%, 3.84% and 6.37% with the various offloading thresholds of 50ms, 100ms and 200ms, respectively. The reason for a low energy saving with various offloading thresholds compared to a high energy saving with the no-offloading approach is that the workload of the primary fog nodes is high, thus sharing some of their workloads with their neighbours. As a result, neighbours staying ON most of the time helps to process these tasks.

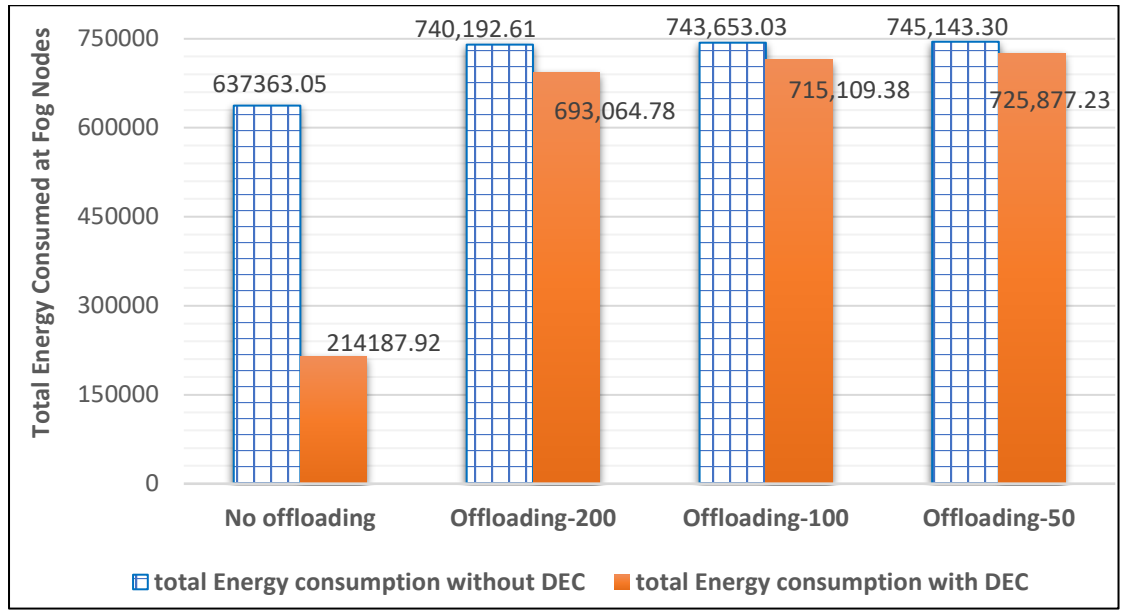

Figure 7: Total Energy Consumed in the fog environment with no offloading and different offloading thresholds with and without Dynamic Energy Controlling (DEC) algorithm

Varying the offloading threshold of the queuing delay, such as 50ms, 100ms and 200ms does improve service latency and throughput. However, this is not the case with energy consumption because more energy is spent by the fog system when the offloading threshold is set to 50ms. This is because more fog nodes are involved in the execution process and therefore require more power to work efficiently. However, after applying the DEC algorithm, energy consumption was reduced. Varying the offloading threshold has not been addressed before in other publications, but it does have a positive impact on overall results. However, this technique depends on the number of neighbouring fog nodes that are willing to help and the availability of these fog nodes. This will be addressed in future work.

## 7. CONCLUSION

In this paper, we studied the problem of minimising service latency and power consumption in fog computing systems and proposed a combination of two efficient and effective algorithms: dynamic task scheduling (DTS) and dynamic energy control (DEC).

In future work, latency and energy overhead caused by activating switched off fog nodes should be considered, and their impact on the system should be addressed. This is because fog nodes are operational devices that require time and energy to boot up in contrast to previous work that powers on switched off datacentres without considering latency and energy overhead. Also, frequent switching ON and OFF of edge devices might lead to edge device failure in the long term and shorten the life of edge devices. Therefore, the frequency of switching fog nodes on and off should be considered and minimised.